\documentclass[amsmath, pra, twocolumn, showpacs, floatfix]{revtex4-1}
\usepackage{graphicx}
\usepackage{dsfont}

\begin{document}   

\title{Chiral force of guided light on an atom}

\author{Fam Le Kien,$^1$ S. Sahar S. Hejazi,$^1$ Viet Giang Truong,$^2$ S\'{i}le Nic Chormaic,$^{2}$ and Thomas Busch$^1$}

\affiliation{$^1$Quantum Systems Unit, Okinawa Institute of Science and Technology Graduate University, Onna, Okinawa 904-0495, Japan\\
$^2$Light-Matter Interactions Unit, Okinawa Institute of Science and Technology Graduate University, Onna, Okinawa 904-0495, Japan
}

\date{\today}

\begin{abstract}
We calculate the force of a near-resonant guided light field of an ultrathin optical fiber on a two-level atom.
We show that, if the atomic dipole rotates in the meridional plane, the magnitude of the force of the guided light depends on the field propagation direction. 
The chirality of the force arises as a consequence of the directional dependencies of the Rabi frequency of the guided driving field and the spontaneous emission from the atom. This provides a unique method for controlling atomic motion in the vicinity of an ultrathin fiber. 
\end{abstract}

\maketitle

Applying controllable optical forces to atoms plays a central role in many areas of physics, in particular in laser cooling and trapping \cite{coolingbook}. 
Various sophisticated schemes for exerting optical forces on atoms have been developed \cite{coolingbook,dipole force}.
A common feature of many cooling and trapping schemes for atoms in free space is that, since spontaneous emission is in a random direction and symmetric with respect to two opposite propagation directions, the average of the recoil over many spontaneous emission events results in a zero net effect on the atomic momentum.

However, it has recently been shown that, for atoms near a nanofiber \cite{Fam2014,Petersen2014,Mitsch14b,sponhigh} or a flat surface \cite{flat}, spontaneous emission may become asymmetric with respect to opposite propagation directions. Such directional spontaneous emission can modify the optical force on atoms. 
In particular, a resonant lateral Casimir-Polder force may arise for an initially excited atom with a rotating dipole near a nanofiber \cite{Scheel15}. 
Such a force appears because, in the presence of a nanofiber, the interaction between the field and the atom with a rotating dipole is chiral \cite{Fam2014,Petersen2014,Mitsch14b,sponhigh}. Chiral optical forces have been studied extensively for chiral molecules and nanoparticles in free space \cite{Ebbesen2013,Doyle2013,Cameron2014,Andrews2014}, in optical lattices \cite{Genet2014}, and near optical nanofibers \cite{Reinhard2016}.
However, under normal conditions, an atom is essentially achiral because it has the high degree of symmetry associated with a sphere \cite{Barron2004}.

The possibility of creating chiral forces acting on atoms holds significant potential in many area of physics, in particular in laser cooling, quantum metrology, and atomic state preparation. 
It enables, for example, the transfer of photonic superposition states to atomic center-of-mass superposition states, opening the possibility of a new way of constructing atomic interferometers. In these, the absorption of a photon superposed in different directions would lead to an atomic motion superposed in the same degree of freedom.
Furthermore, directional forces could help with sorting atoms to achieve optical lattices with unit filling factors \cite{Barredo:16,Endres:16}, or lead to new laser cooling schemes that can exceed the recoil limit.

In this Letter, we calculate the force of a near-resonant guided light field of an ultrathin optical fiber on a two-level atom.
We show that directional absorption and emission of guided photons leads to a significant chiral optical force.
    
\begin{figure}[tbh]
\begin{center}
  \includegraphics{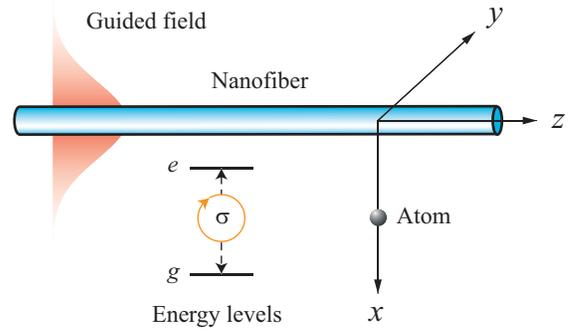}
 \end{center}
\caption{(Color online) A two-level atom with a dipole rotating in the meridional plane $zx$ outside an ultrathin optical fiber. 
}
\label{fig1}
\end{figure}

We study a two-level atom driven by a near-resonant classical field with optical frequency $\omega_L$ and slowly varying envelope $\boldsymbol{\mathcal{E}}$ near a vacuum-clad ultrathin optical fiber (see Fig.~\ref{fig1}). 
The atom has an upper energy level $|e\rangle$ and a lower energy level $|g\rangle$, with energies $\hbar\omega_e$ and $\hbar\omega_g$, respectively. 
The atomic transition frequency is $\omega_0=\omega_e-\omega_g$. The fiber is a dielectric cylinder of radius $a$ and refractive index $n_1>1$ and is surrounded by an infinite background vacuum or air medium of refractive index $n_2$, where $n_2=1$. 
For the atomic position and the field distribution,
we use Cartesian coordinates $\{x,y,z\}$, where $z$ is the coordinate along the fiber axis, and also cylindrical coordinates $\{r,\varphi,z\}$, where $r$ and $\varphi$ are the polar coordinates in the fiber transverse plane $xy$.
 
The atom interacts with the classical driving field $\boldsymbol{\mathcal{E}}$ and the quantum electromagnetic field via the spontaneous emission process. The positive-frequency part $\mathbf{E}$ of the electric component of the quantum field can be
decomposed into the contributions from the guided modes, $\mathbf{E}_{\mathrm{g}}$, and the radiation modes, $\mathbf{E}_{\mathrm{r}}$.
In view of the very low losses of silica in the wavelength range of interest, we neglect material absorption. 

The Hamiltonian for the atom-field interaction in the dipole approximation is given by 
\begin{eqnarray}\label{c13}
H_{\mathrm{int}}&=&-\frac{\hbar}{2}\Omega\sigma_{eg}e^{-i(\omega_L-\omega_0)t}
-i\hbar\sum_{\alpha}G_{\alpha}\sigma_{eg} a_{\alpha}e^{-i(\omega-\omega_0)t}\nonumber\\
&&\mbox{}
-i\hbar\sum_{\alpha}\tilde{G}_{\alpha}\sigma_{ge} a_{\alpha}e^{-i(\omega+\omega_0)t}
+\mbox{H.c.},
\end{eqnarray}
where $\sigma_{ij}=|i\rangle\langle j|$ with $i,j=e,g$ are the atomic operators, $a_{\alpha}$ and $a_{\alpha}^\dagger$ are the photon operators, 
$\Omega=\mathbf{d}_{eg}\cdot\boldsymbol{\mathcal{E}}/\hbar$ is the Rabi frequency produced by the driving field,
with $\mathbf{d}_{eg}=\langle e|\mathbf{D}|g\rangle$ being the matrix element of the atomic dipole operator $\mathbf{D}$,
and $G_\alpha$ and $\tilde{G}_\alpha$ are the coupling coefficients for the interaction between the atom and the quantum field.
The notations $\alpha=\mu,\nu$ and $\sum_{\alpha}=\sum_{\mu}+\sum_{\nu}$ stand for the general mode index and the mode summation, respectively.
The index $\mu=(\omega N f p)$ labels guided modes, where $\omega$ is the mode frequency, $N=\mathrm{HE}_{lm}$, EH$_{lm}$, TE$_{0m}$, or TM$_{0m}$ is the mode type, with $l=1,2,\dots$ and $m=1,2,\dots$ being the azimuthal and radial mode orders, respectively, $f=\pm1$ denotes the forward or backward propagation direction along the fiber axis $z$, and $p$ is the polarization index.  
The index $\nu=(\omega \beta l p)$ labels radiation modes, where $\beta$ is the longitudinal propagation constant,  
$l=0,\pm1,\pm2,\dots$ is the mode order, and $p=+,-$ is the mode polarization.
The notations $\sum_{\mu}=\sum_{N fp}\int_0^{\infty}d\omega$ and $\sum_{\nu}=\sum_{lp}\int_0^{\infty}d\omega\int_{-kn_2}^{kn_2}d\beta$
are the generalized summations over the guided and radiation modes, respectively.   
 
The expressions for the coupling coefficients $G_{\alpha}$ and $\tilde{G}_{\alpha}$ with $\alpha=\mu,\nu$ are given as  
\begin{eqnarray}\label{c14}
G_{\mu}&=&\sqrt{\frac{\omega\beta'}{4\pi\epsilon_0\hbar}}\;
(\mathbf{d}_{eg}\cdot\mathbf{e}^{(\mu)})e^{i(f\beta z+pl\varphi)},\nonumber\\
G_{\nu}&=&\sqrt{\frac{\omega}{4\pi\epsilon_0\hbar}}\;
(\mathbf{d}_{eg}\cdot\mathbf{e}^{(\nu)})e^{i(\beta z+l\varphi)},
\end{eqnarray}
and
\begin{equation}\label{c15}
\begin{split}
\tilde{G}_{\mu}&=\sqrt{\frac{\omega\beta'}{4\pi\hbar\epsilon_0}}\;
(\mathbf{d}^*_{eg}\cdot\mathbf{e}^{(\mu)})e^{i(f\beta z+pl\varphi)},\\
\tilde{G}_{\nu}&=\sqrt{\frac{\omega}{4\pi\hbar\epsilon_0}}\;
(\mathbf{d}^*_{eg}\cdot\mathbf{e}^{(\nu)})e^{i(\beta z+l\varphi)},
\end{split}
\end{equation}
where $\mathbf{e}^{(\mu)}$ and $\mathbf{e}^{(\nu)}$ are the normalized mode functions given in Refs. \cite{fiber books,sponhigh}.
An important property of the mode functions of hybrid HE and EH modes and TM modes is that the longitudinal
component $e_z^{(\mu)}$ is nonvanishing and in quadrature ($\pi/2$ out of phase) with the radial component $e_r^{(\mu)}$.
We note that in deriving the Hamiltonian (\ref{c13}) we have used the rotating-wave approximation for the driving field $\boldsymbol{\mathcal{E}}$. 
 
In a semiclassical treatment, the center-of-mass motion of the atom is governed by the force that is defined by the formula
$\mathbf{F}= -\langle\boldsymbol{\nabla} H_{\mathrm{int}}\rangle$ \cite{coolingbook,dipole force}.
In the framework of the Born-Markov approximation, we find
\begin{equation}\label{c25}
\mathbf{F}=\mathbf{F}_{\mathrm{drv}}+\rho_{ee}\mathbf{F}_{\mathrm{spon}}+\rho_{ee}\mathbf{F}_{\mathrm{vdW}}^{(e)}
+\rho_{gg}\mathbf{F}_{\mathrm{vdW}}^{(g)},             
\end{equation}
where
\begin{eqnarray}\label{c26}
\mathbf{F}_{\mathrm{drv}}&=&\frac{\hbar}{2}(\rho_{ge}\boldsymbol{\nabla}\Omega+\rho_{eg}\boldsymbol{\nabla}\Omega^*),\nonumber\\
\mathbf{F}_{\mathrm{spon}}&=&i\pi\hbar\sum_{\alpha_0}(G_{\alpha_0}^*\boldsymbol{\nabla}G_{\alpha_0}-G_{\alpha_0}\boldsymbol{\nabla}G_{\alpha_0}^*),\nonumber\\
\mathbf{F}_{\mathrm{vdW}}^{(e)}&=&\hbar\boldsymbol{\nabla}\mathcal{P}\sum_{\alpha}\frac{|G_{\alpha}|^2}{\omega-\omega_0},\quad
\mathbf{F}_{\mathrm{vdW}}^{(g)}=\hbar\boldsymbol{\nabla}\mathcal{P}\sum_{\alpha}\frac{|\tilde{G}_{\alpha}|^2}{\omega+\omega_0}. \qquad            
\end{eqnarray}
Here, $\mathbf{F}_{\mathrm{drv}}$ is the force produced by the interaction between the driving field and the atom (the recoil of absorption and the dynamical Stark shifts of the energy levels),
$\mathbf{F}_{\mathrm{spon}}$ is the force produced by spontaneous emission from the excited state, and
$\mathbf{F}_{\mathrm{vdW}}^{(e)}$ and $\mathbf{F}_{\mathrm{vdW}}^{(g)}$ 
are the forces associated with the surface-induced van der Waals potentials for the excited and ground states, respectively.
The notation $\rho$ stands for the density operator of the internal atomic state in the coordinate frame rotating with the frequency $\omega_L$,
the notation $\alpha_0=\mu_0,\nu_0$ labels resonant guided modes $\mu_0=(\omega_0 N f p)$ or resonant radiation modes $\nu_0=(\omega_0 \beta l p)$, the generalized summation $\sum_{\alpha_0}$ is $\sum_{\alpha_0}=\sum_{\mu_0}+\sum_{\nu_0}$ with
$\sum_{\mu_0}=\sum_{N fp}$ and $\sum_{\nu_0}=\sum_{lp}\int_{-k_0n_2}^{k_0n_2}d\beta$, and the notation $\mathcal{P}$ stands for the principal value of the integral over $\omega$. 

The forces $\mathbf{F}_{\mathrm{vdW}}^{(e)}$ and $\mathbf{F}_{\mathrm{vdW}}^{(g)}$ result from the surface-induced potentials 
$U_e=-\hbar\mathcal{P}\sum_{\alpha}|G_{\alpha}|^2/(\omega-\omega_0)-\delta E_e^{(\mathrm{vac})}$ and 
$U_g=-\hbar\mathcal{P}\sum_{\alpha}|\tilde{G}_{\alpha}|^2/(\omega+\omega_0)-\delta E_g^{(\mathrm{vac})}$,  
where $\delta E_e^{(\mathrm{vac})}$ and $\delta E_g^{(\mathrm{vac})}$ are the shifts of the energy levels induced by the vacuum field in free space.
Note that $\delta E_e^{(\mathrm{vac})}-\delta E_g^{(\mathrm{vac})}=\hbar\delta\omega_0^{(\mathrm{vac})}$, where $\delta\omega_0^{(\mathrm{vac})}$ is the Lamb shift.
The environment-induced shift of atomic transition frequency is $\delta\omega_0=\delta\omega_0^{(\mathrm{vac})}+(U_e-U_g)/\hbar$.
The shifted atomic transition frequency is $\omega_{A}=\omega_0+\delta\omega_0$.
When the atom is not too close to the fiber, we have $|\delta\omega_0|\ll\omega_0$, 
which leads to  $\omega_{A}\simeq\omega_0$. 
We formally incorporate $\delta\omega_0$ into $\omega_0$.

Equation \eqref{c25} is valid for an arbitrary driving field, which includes the incident field and the scattered field.
When the atom is in free space, we have $\mathbf{F}_{\mathrm{spon}}=\mathbf{F}_{\mathrm{vdW}}^{(e)}=\mathbf{F}_{\mathrm{vdW}}^{(g)}=0$, which leads to
$\mathbf{F}=\mathbf{F}_{\mathrm{drv}}$, that is, the force on the atom is just the conventional radiation force \cite{coolingbook,dipole force}.
We note that $\mathbf{F}_{\mathrm{vdW}}^{(g)}$ and $\mathbf{F}_{\mathrm{spon}}+\mathbf{F}_{\mathrm{vdW}}^{(e)}$
are the total surface-induced forces for the ground and excited states, respectively. These forces have previously been calculated 
using the Green function approach \cite{Buhmann2004}. When the excitation of the atom is weak, our results reduce to those
of Ref.~\cite{Dogariu2016} for a point dipole near an interface.  
 
We now assume that the driving field is in a guided mode 
propagating along the fiber axis $z$ with the propagation constant $\beta_L$ in the $f_L$ direction,
that is, $\boldsymbol{\mathcal{E}}=\boldsymbol{\mathcal{E}}_0(r,\varphi)e^{if_L\beta_Lz}$. 
We are interested in the axial component $F_z$ of the force. Due to the symmetry of the system,
the potentials $U_e$ and $U_g$ do not depend on $z$.  
Therefore, we find
\begin{eqnarray}\label{c35}
F_z&=&\frac{i\hbar f_L\beta_L}{2}(\Omega\rho_{ge}-\Omega^*\rho_{eg})
-\rho_{ee}\hbar\sum_{N}\beta_{0}^{(N)}(\gamma_{\mathrm{g}N}^{(+)}-\gamma_{\mathrm{g}N}^{(-)})
\nonumber\\&&\mbox{}
-\rho_{ee}\int_{-k_0n_2}^{k_0n_2}\hbar\beta\gamma^{(\beta)}_{\mathrm{r}} d\beta,
\end{eqnarray}
where $\beta_{0}^{(N)}$ is the propagation constant of the guided modes $N$ at the frequency $\omega_0$,
$\gamma_{\mathrm{g}N}^{(f)}=2\pi \sum_{p}|G_{\omega_0Nfp}|^2$
is the rate of spontaneous emission into the guided modes $N$ with the propagation direction $f$ and
$\gamma^{(\beta)}_{\mathrm{r}}=2\pi\sum_{lp}|G_{\omega_0\beta lp}|^2$
is the rate of spontaneous emission into the radiation modes with the axial wave-vector component $\beta$ \cite{sponhigh}. 
Note that the first term in Eq.~(\ref{c35}) is the recoil of the absorption, while the second and third terms are the recoils of spontaneous emission into guided and radiation modes.

To calculate the axial force $F_z$ in detail, we first assume that the atom is at rest and in the steady state.
We can then use the steady-state solution for the internal state of the atom and find  
\begin{equation}\label{c42}
F_z=\hbar\rho_{ee}\bigg\{f_L\beta_L\Gamma
-\sum_{N}\beta_0^{(N)}(\gamma_{\mathrm{g}N}^{(+)}-\gamma_{\mathrm{g}N}^{(-)})
-\int\limits_{-k_0n_2}^{k_0n_2}\beta\gamma^{(\beta)}_{\mathrm{r}} d\beta\bigg\},
\end{equation}
where
\begin{equation}\label{c40}
\rho_{ee}=\frac{|\Omega|^2/4}{\Delta^2+\Gamma^2/4+|\Omega|^2/2}.
\end{equation}
Here, $\Delta=\omega_L-\omega_0$ is the detuning of the driving-field frequency $\omega_L$ 
from the  atomic transition frequency $\omega_0$, and
$\Gamma=\gamma_{\mathrm{g}}+\gamma_{\mathrm{r}}$
is the rate of spontaneous emission, with
$\gamma_{\mathrm{g}}=\sum_{N}(\gamma_{\mathrm{g}N}^{(+)}+\gamma_{\mathrm{g}N}^{(-)})$
being the rate of emission into guided modes and
$\gamma_{\mathrm{r}}=\int_{-k_0n_2}^{k_0n_2}\gamma^{(\beta)}_{\mathrm{r}} d\beta$
being the rate of emission into radiation modes \cite{sponhigh}.

For an atom with a circular dipole near a nanofiber, the spontaneous emission rates $\gamma_{\mathrm{g}N}^{(f)}$ and $\gamma^{(\beta)}_{\mathrm{r}}$ can be asymmetric with respect to the opposite axial propagation directions \cite{Fam2014,Petersen2014,Mitsch14b,flat,sponhigh,Scheel15}. These directional effects are the signatures of spin-orbit coupling of light carrying transverse spin angular momentum \cite{Zeldovich,Bliokh review,Bliokh review2015,Bliokh2014,Bliokh2015,Lodahl2017,Banzer review2015}. They are due to the existence of a nonzero longitudinal component of the field in the presence of the nanofiber. This component oscillates in phase quadrature with respect to the radial transverse component and, hence, makes the field chiral.
The effect occurs when the atom has a dipole rotating in the meridional plane, that is, when the atom is chiral and the ellipticity vector of the dipole
overlaps with the ellipticity vector of the field  \cite{Fam2014,Petersen2014,Mitsch14b,flat,sponhigh,Scheel15}.
As a consequence, the absolute value of the force $F_z$, given by Eq.~\eqref{c42}, can also be asymmetric with respect to the opposite propagation directions 
$f_L=\pm$ of the driving field. The asymmetry of the force can be characterized by the parameter $\eta=(|F_z^{(+)}|-|F_z^{(-)}|)/(|F_z^{(+)}|+|F_z^{(-)}|)$. Here, $F_z^{(\pm)}$ is the force when the driving field propagates in the direction $\pm z$. 

We now calculate numerically the dependence of the magnitude of the force $F_z$ on the propagation direction $f_L$ of the driving field.
We assume that the atom is positioned on the $x$ axis and the dipole matrix element $\mathbf{d}$ is a complex vector in the meridional plane $zx$
(see Fig.~\ref{fig1}). To be concrete, we take $\mathbf{d}_{eg}=d(i\hat{\mathbf{x}}-\hat{\mathbf{z}})/\sqrt2$, 
which corresponds to the $\sigma^+$ transition with respect to the quantization $y$ axis. The results for the $\sigma^-$ transition can be obtained
from the results for the $\sigma^+$ by replacing $F_z^{(+)}$ and $F_z^{(-)}$ with $-F_z^{(-)}$ and $-F_z^{(+)}$, respectively.

We assume that the driving field is prepared in a quasilinearly polarized hybrid HE or EH mode or a TM mode.
In the case of HE and EH modes, we choose the $x$ polarization, which leads to a maximal longitudinal component of the field at the position of the atom.
We do not consider the case of a TE mode because of the vanishing of the interaction between such a mode and the chosen atomic dipole. 
For an $x$-polarized hybrid HE or EH mode or a TM mode with the propagation direction $f_L$, the field amplitude at the position of the atom is
$\boldsymbol{\mathcal{E}}(r,\varphi=0,z=0)=\mathcal{A}(e_r\hat{\mathbf{x}}+f_Le_z\hat{\mathbf{z}})$, where $\mathcal{A}$ is determined by the power of the driving field \cite{fiber books,highorder}. The corresponding Rabi frequency is $\Omega=(d\mathcal{A}/\hbar\sqrt2)(ie_r-f_Le_z)$.
Since the relative phase between the complex functions $e_r$ and $e_z$ is $\pi/2$ \cite{fiber books,highorder}, the absolute value $|\Omega|$ of the Rabi frequency
depends on the propagation direction $f_L$. This leads to a direction dependence of the excited-state population $\rho_{ee}$ and, hence, contributes
to the asymmetry of the force $F_z$. Thus, both excitation and spontaneous emission can contribute to the dependence of the force $F_z$ on the propagation direction of the driving field. Note that the effects of excitation and spontaneous emission on the asymmetry of the force $F_z$ may enhance or partially
compensate each other.

\begin{figure}[tbh]
\begin{center}
 \includegraphics{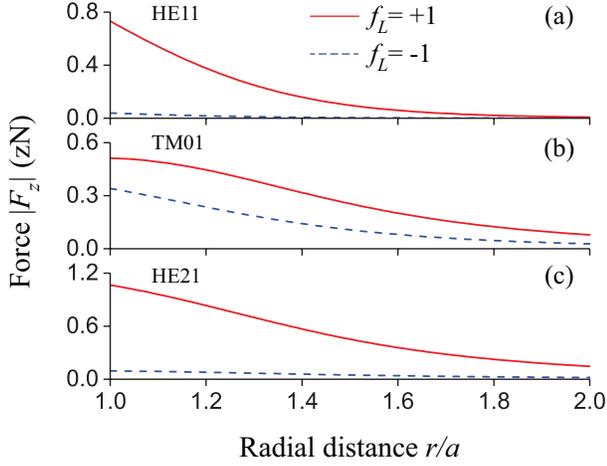}
 \end{center}
\caption{(Color online) Radial dependence of the absolute value $|F_z|$ of the force of the resonant guided light on a two-level atom. The incident light field is 
in (a) a $x$-polarized HE$_{11}$ mode, (b) a TM$_{01}$ mode, or (c) a $x$-polarized HE$_{21}$ mode and propagates in the forward $f_L=+$ (solid red curves) 
or backward $f_L=-$ (dashed blue curves) direction along the fiber axis $z$ with the power $P=1$ pW.
The dipole matrix element of the atom is $\mathbf{d}=d(i\hat{\mathbf{x}}-\hat{\mathbf{z}})/\sqrt2$,  
the fiber radius is $a=350$ nm, and the wavelength of the atomic transition is $\lambda_0=780$ nm. 
The refractive indices of the fiber and the vacuum cladding are $n_1=1.4537$ and $n_2=1$, respectively.}
\label{fig2}
\end{figure}

The radial dependencies of the absolute value $|F_z|$ of the force for the cases where the driving field is in an   
$x$-polarized HE$_{11}$ mode, a TM$_{01}$ mode, or a $x$-polarized HE$_{21}$ mode with the propagation direction $f_L=\pm$ are shown in Fig.~\ref{fig2}. 
One can see that the absolute value $|F_z|$ of the force has different magnitudes for different propagation directions $f_L$ of the driving field.
This chiral effect occurs not only for the fundamental mode HE$_{11}$ but also for higher-order hybrid HE and EH modes and TM modes.
Figure \ref{fig2}(a) shows that the force of the HE$_{11}$ mode on the atom is almost fully chiral.

\begin{figure}[tbh]
\begin{center}
 \includegraphics{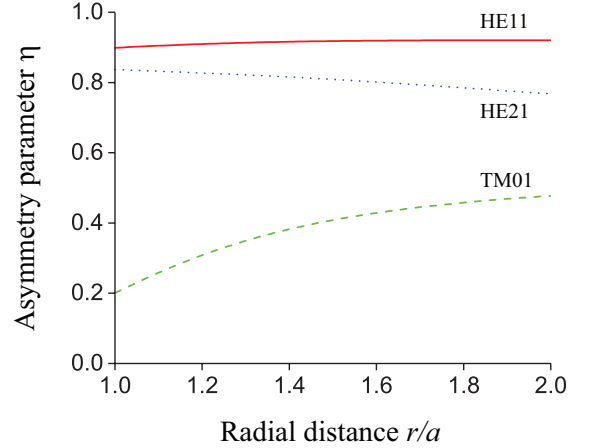}
 \end{center}
\caption{(Color online)  Radial dependence of the asymmetry parameter $\eta$ for the forces $F_z^{(f_L)}$ for the opposite propagation directions $f_L=\pm$. 
The parameters used are the same as for Fig.~\ref{fig2}.}
\label{fig3}
\end{figure}

While the absolute value $|F_z|$ of the force reduces quickly with increasing radial distance $r$, the asymmetry parameter $\eta$ can be seen in Fig.~\ref{fig3} to vary slowly. Moreover, in the limit of large distances, $\eta$ approaches a nonzero limiting value. 
This result means that, despite the evanescent wave behavior of the force, the asymmetry parameter $\eta$ can be significant even when the atom is far away from the fiber. The reason is that $\eta$ is determined by not the field amplitude but the ratio between the axial and radial components of the guided field. 
Indeed, in the limit of large $r$, we have $|F_z|\propto \rho_{ee}\propto |\Omega|^2$. 
This leads to $\eta\simeq 2\mathrm{Im}(e_re_z^*)/(|e_r|^2+|e_z|^2)$ for $\mathbf{d}\propto i\hat{\mathbf{x}}-\hat{\mathbf{z}}$ and 
$\boldsymbol{\mathcal{E}}\propto e_r\hat{\mathbf{x}}+f_Le_z\hat{\mathbf{z}}$.
We can show that $e_z/e_r\to -iq_L/\beta_L$  for $r\to\infty$, where $q_L$ is the evanescent-wave penetration parameter for the driving field \cite{fiber books,highorder}. Hence, we find $\eta\to\eta_{\infty}=2\beta_L q_L/(\beta_L^2+q_L^2)$ for $r\to\infty$.
Thus, the limiting value of $\eta$ is nonzero and is determined by the fiber guided mode parameters $\beta_L$ and $q_L$. 
Since $q_L=\sqrt{\beta_L^2-n_2^2k_L^2}<\beta_L\leq n_1 k_L$, we have $\eta_{\infty}\le 2n_1\sqrt{n_1^2-n_2^2}/(2n_1^2-n_2^2)<1$.
It is clear that one can enhance the limiting value $\eta_{\infty}$ by increasing the refractive index $n_1$ of the fiber. 

It is interesting to note that the asymptotic value $\eta\simeq 2\mathrm{Im}(e_re_z^*)/(|e_r|^2+|e_z|^2)$ for large $r$ is proportional to the electric transverse spin density $\rho_y^{\text{e-spin}}\equiv (\epsilon_0/4\omega_L) 
\mathrm{Im}[\boldsymbol{\mathcal{E}}^*\times\boldsymbol{\mathcal{E}}]\cdot\hat{\mathbf{y}}
\propto f_L\mathrm{Im}(e_re_z^*)$ of the guided driving field \cite{highorder}. A simple explanation is that, for the atom with a dipole rotating in the meridional plane $zx$, the axis $y$ is the quantization axis and, hence, the selection rule corresponds to the transverse spin angular momentum conservation. Due to this fact, the Rabi frequency is determined by the field spherical tensor component which rotates
in the same direction as that of the dipole in the plane $zx$. When the propagation direction of light is reversed, the rotation direction of the spin angular momentum is also reversed in accordance with the spin-orbit coupling of light  \cite{Zeldovich,Bliokh review,Bliokh review2015,Bliokh2014,Bliokh2015,Lodahl2017,Banzer review2015}.
Therefore, the difference between the squared absolute values of the Rabi frequencies for the opposite propagation
directions is proportional to the difference between the squared absolute values of the opposite spherical tensor components of the guided driving field in the plane $zx$. Meanwhile, the first difference is proportional to the difference between the excitations and, hence, to the asymptotic difference between the forces for large $r$, and the second difference is proportional to the electric transverse spin density. This explains why the asymptotic value of the asymmetry parameter $\eta$ is proportional to the electric transverse spin density $\rho_y^{\text{e-spin}}$. 
Thus, the asymmetry of the forces is a signature of spin-orbit coupling of light \cite{Zeldovich,Bliokh review,Bliokh review2015,Bliokh2014,Bliokh2015,Lodahl2017,Banzer review2015}.

\begin{figure}[tbh]
\begin{center}
 \includegraphics{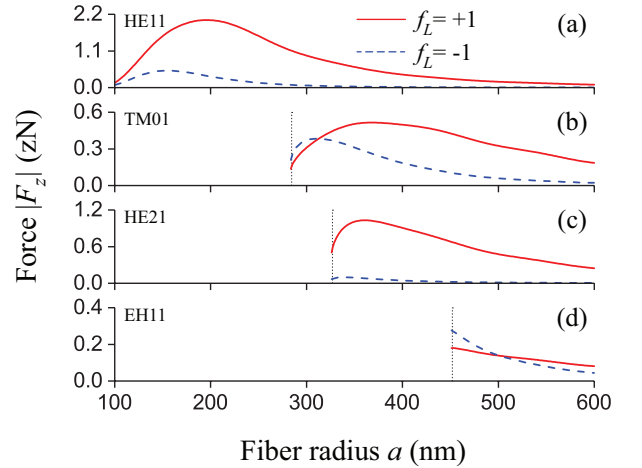}
 \end{center}
\caption{(Color online) Absolute value $|F_z|$ of the force of the resonant guided light on a two-level atom as a function of the fiber radius $a$. The atom is positioned at the distance
$r-a=20$ nm from the fiber surface. Other parameters are as for Fig.~\ref{fig2}. The vertical dotted lines indicate the positions of the cutoffs for higher-order modes.}
\label{fig4}
\end{figure}

\begin{figure}[tbh]
\begin{center}
 \includegraphics{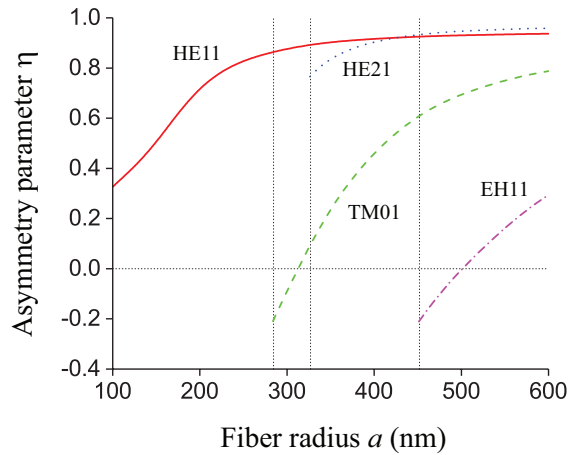}
 \end{center}
\caption{(Color online) Asymmetry parameter $\eta$ for the forces $F_z^{(f_L)}$ for the opposite propagation directions $f_L=\pm$ as a function of the fiber radius $a$. The parameters used are the same as for Fig.~\ref{fig4}. The vertical dotted lines indicate the positions of the cutoffs for higher-order modes.}
\label{fig5}
\end{figure}

The dependencies of the absolute value $|F_z|$ and the directional asymmetry parameter $\eta$ of the force on the fiber radius $a$ are shown in Figs.~\ref{fig4} and \ref{fig5}. While for the modes HE$_{11}$ and HE$_{21}$ one always find $|F_z^{(+)}|\geq|F_z^{(-)}|$,
the modes TM$_{01}$ and EH$_{11}$ allow for both possibilities $|F_z^{(+)}|\leq|F_z^{(-)}|$ and $|F_z^{(+)}|\geq|F_z^{(-)}|$, depending on $a$.

In summary, we have calculated the force of the guided light field of an ultrathin optical fiber on a two-level atom.
We have shown that the magnitude of the force of guided light on an atom with a dipole rotating in the meridional plane depends on the field propagation direction. This chiral effect arises as a consequence of the directional dependencies of the Rabi frequency of the guided driving field and the spontaneous emission from the atom. Our results could be used to control and manipulate the direction of motion of atoms in a cold gas or an optical lattice
near the surface of an ultrathin fiber by simply varying the field propagation direction. This could enable studies of optical binding effects on atoms under chiral forces, and lead to new laser cooling schemes and novel designs for atom interferometers.  

We acknowledge support for this work from the Okinawa Institute of Science and Technology Graduate University.


\end{document}